\documentclass[useAMS,usenatbib]{mn2e}
\usepackage{graphicx}
\usepackage{times}

\newcommand{\Ion}[2]{#1{\,\sc#2}}
\newcommand{\kms}{\mbox{$\mathrm{km\,s^{-1}}$}}
\newcommand{\MSUN}{\mbox{$\mathrm{M_{\odot}}$}}
\newcommand{\RSUN}{\mbox{$\mathrm{R_{\odot}}$}}

\title[An eclipsing magnetic white dwarf]{A magnetic white dwarf in a detached eclipsing binary}

\author[S. G. Parsons et al.]{S.~G.~Parsons$^{1}$\thanks{steven.parsons@uv.cl},
T.~R.~Marsh$^{2}$,
B.~T.~G{\"a}nsicke$^{2}$,
M.~R.~Schreiber$^{1}$,
M.~C.~P.~Bours$^{2}$,
\newauthor
V.~S.~Dhillon$^{3}$,
and S.~P.~Littlefair$^{3}$
\\
$^{1}$ Departmento de F{\'i}sica y Astronom{\'i}a, Universidad de
Valpara{\'i}so, Avenida Gran Bretana 1111, Valpara{\'i}so, Chile\\
$^{2}$ Department of Physics, University of Warwick, Coventry CV4 7AL, UK\\
$^{3}$ Department of Physics and Astronomy, University of Sheffield,
Sheffield, S3 7RH, UK}
\begin{document}
\input{references.cls}
\date{Accepted 2013 August 16.  Received 2013 August 16; in original form 2013 July 15}

\pagerange{\pageref{firstpage}--\pageref{lastpage}} \pubyear{2013}

\maketitle

\label{firstpage}

\begin{abstract}

SDSS\,J030308.35+005444.1 is a close, detached, eclipsing white dwarf plus M dwarf binary which shows a large infrared excess which has been interpreted in terms of a circumbinary dust disk. In this paper we present optical and near-infrared photometric and spectroscopic data for this system. At optical wavelengths we observe heated pole caps from the white dwarf caused by accretion of wind material from the main-sequence star on to the white dwarf. At near-infrared wavelengths we see the eclipse of two poles on the surface of the white dwarf by the main-sequence star, indicating that the white dwarf is magnetic. Our spectroscopic observations reveal Zeeman split emission lines in the hydrogen Balmer series, which we use to measure the magnetic field strength as 8\,MG. This measurement indicates that the cyclotron lines are located in the infrared, naturally explaining the infrared excess without the need for a circumbinary dust disk. We also detect magnetically-confined material located roughly midway between the two stars. Using measurements of the radial velocity amplitude and rotational broadening of the M star we constrain the physical parameters of the system, a first for a magnetic white dwarf, and the location of the poles on the surface of the white dwarf. SDSS\,J030308.35+005444.1 is a pre-cataclysmic variable that will likely evolve into an intermediate polar in $\sim$1\,Gyr.

\end{abstract}


\section{Introduction}

Cataclysmic variables (CVs) are semi-detached binaries containing a white dwarf accreting material from a main-sequence companion star via Roche-lobe overflow. Around 20\% of CVs contain white dwarfs with magnetic fields strong enough to effect the flow of material on to it \citep{wickramasinghe00,araujo05,pretorius13}. These magnetic CVs are divided into two types: polars, in which the white dwarf is synchronously rotating with the binary orbit, and intermediate polars (IPs), in which the white dwarfs usually rotate faster than the binary orbit.

Magnetic CVs are thought to have a similar evolutionary history to non-magnetic CVs. Both initially started out as main-sequence binaries that underwent a common envelope phase, where the two stars orbited within a single envelope of material, losing angular momentum and spiraling closer together. The resulting system consists of a close, detached white dwarf plus main-sequence binary, known as a post common envelope binary (PCEB). These systems have orbital periods of a few hours to a few days and gradually lose angular momentum via gravitational radiation (\citealt{kraft62}; \citealt{faulkner71}) and magnetic braking (\citealt{verbunt81}; \citealt{rappaport83}) eventually bringing them close enough together to initiate mass transfer on to the white dwarf.

\begin{table*}
 \centering
  \caption{Journal of observations. Exposure times for X-shooter observations are for UVB arm, VIS arm and NIR arm respectively. The primary eclipse occurs at phase 1, 2 etc.}
  \label{tab:obs_log}
  \begin{tabular}{@{}lccccccc@{}}
  \hline
  Date at     &Instrument&Telescope & Filter(s)&Start  & Orbital  &Exposure   &Conditions                 \\
  start of run&          &          &          &(UT)   & phase    &time (s)   &(Transparency, seeing)     \\
  \hline
  2007/10/17  & ULTRACAM & WHT      & $u'g'i'$ &02:34  &0.88--1.18&5.2        & Good, $\sim$1.2 arcsec    \\
  2007/10/18  & ULTRACAM & WHT      & $u'g'i'$ &02:25  &0.28--1.52&5.2        & Good, $\sim$1.2 arcsec    \\
  2007/10/29  & ULTRACAM & WHT      & $u'g'i'$ &04:40  &0.80--1.09&2.3        & Poor, $>$3 arcsec         \\
  2010/11/10  & ULTRACAM & NTT      & $u'g'i'$ &04:27  &0.47--1.25&4.3        & Excellent, $<$1 arcsec    \\ 
  2010/11/11  & ULTRACAM & NTT      & $u'g'i'$ &03:16  &0.54--1.80&4.3        & Good, $\sim$1.2 arcsec    \\
  2010/11/25  & ULTRACAM & NTT      & $u'g'i'$ &03:18  &0.69--1.16&3.8        & Excellent, $<$1 arcsec    \\ 
  2012/09/09  & ULTRACAM & WHT      & $u'g'r'$ &03:18  &0.80--1.70&4.0        & Good, $\sim$1.2 arcsec    \\
  2012/10/09  & ULTRACAM & WHT      & $u'g'r'$ &23:51  &0.91--1.12&5.0        & Good, $\sim$1.4 arcsec    \\
  2012/10/13  & ULTRACAM & WHT      & $u'g'r'$ &00:33  &0.89--1.11&3.8        & Excellent, $<$1 arcsec    \\ 
  2012/11/16  & X-shooter& VLT      & -        &00:57  &0.47--2.25&606,294,100& Excellent, $<$1 arcsec    \\ 
  2012/11/30  & HAWK-I   & VLT      & $K_\mathrm{S}$    &00:31  &0.48--1.69&1.0        & Average, $\sim$1.5 arcsec \\
  \hline
\end{tabular}
\end{table*}

However, whilst the number of identified PCEBs has increased rapidly in the last few years, there is a complete lack of systems containing obviously magnetic white dwarfs \citep{liebert05,silvestri06,silvestri07,rebassa10,rebassa12}. Given that at least 10\% of isolated white dwarfs also show evidence of magnetic fields with strengths in excess of 2\,MG \citep{liebert03}, this is a large discrepancy. 

Magnetic white dwarfs are thought to be generally more massive (hence smaller and fainter) than non-magnetic white dwarfs \citep{kulebi10,dobbie12,dobbie13,kulebi13}, which could account for some of this discrepancy, but cannot explain the complete absence of strongly Zeeman-split hydrogen lines in the more than 2000 spectroscopically identified white dwarf plus main-sequence binaries \citep{rebassa12}.

\citet{ferrario12} recently showed that the initial mass ratio distribution implies that there should exist a large population of close, detached white dwarf plus F or G type main-sequence stars. These are descendents of somewhat more massive main-sequence binaries and so could harbour magnetic white dwarfs, invisible at optical wavelengths due to their bright main-sequence companions. An alternative hypothesis involves the generation of magnetic field during the common envelope phase itself via a magnetic dynamo \citep{tout08,potter10}. In this scenario, systems with strongly magnetic white dwarfs emerge from the common envelope at small separations and hence evolve into CVs on short timescales.

There are a small number of magnetic white dwarfs accreting from the wind of their main-sequence companion stars \citep{reimers99,reimers00,schmidt05,schmidt07,vogel07,vogel11,schwope09}. These systems are characterised by cool white dwarfs with field strengths of a few 10\,MG (i.e. cyclotron lines in the optical) and main-sequence stars that underfill their Roche-lobes, hence these systems are in fact pre-polars, yet to become CVs. The accretion rates in these systems are of the order of $10^{-13}$--$10^{-14}$\,{\MSUN}/yr, several orders of magnitude smaller than in normal polars, but much higher than in equivalent detached non-magnetic PCEBs \citep{debes06,tappert11,pyrzas12,parsons12gk,ribeiro13} which have accretion rates of the order of $10^{-17}$--$10^{-15}$\,{\MSUN}/yr. The exception to this is the non-magnetic system QS\,Vir which has a mass transfer rate of $10^{-13}$\,{\MSUN}/yr \citep{matranga12}, although this system may be a hibernating CV \citep{odonoghue03,matranga12}.

SDSS\,J030308.35+005444.1 (henceforth SDSS J0303+0054) was identified as an eclipsing PCEB containing a cool ($\sim$8000\,K) featureless DC white dwarf with a low-mass M dwarf companion in a 3.2\,hour orbit by \citet{pyrzas09}. \citet{parsons10} presented high-speed photometry of the system and noted an unusual modulation in the $u'$ and $g'$ bands that could not be attributed to the ellipsoidal distortion of the M star, they proposed that starspots on the M star could be responsible. Recently, \citet{debes12} discovered a large infrared excess in the system which they argue is caused by a circumbinary dust disk.

In this paper we present new optical and near-infrared photometric and spectroscopic data for SDSS\,J0303+0054 which shows that the white dwarf is in fact magnetic, eliminating the need for a highly spotted M star or circumbinary disk. We also constrain the physical parameters and evolution of the binary. 

\section{Observations and their reduction}

\subsection{ULTRACAM photometry}

SDSS\,J0303+0054 was observed with ULTRACAM mounted as a visitor instrument on the $4.2$m William Herschel Telescope (WHT) on La Palma in 2007 and 2012 and at the $3.5$m New Technology Telescope (NTT) on La Silla in 2010. ULTRACAM is a high-speed, triple-beam CCD camera \citep{dhillon07} which can acquire simultaneous images in three different bands; for our observations we used the SDSS $u'$, $g'$ and either $r'$ or $i'$ filters. A complete log of these observations is given in Table~\ref{tab:obs_log}. We windowed the CCD in order to achieve exposure times of $\sim$3--4 seconds which we varied to account for the conditions; the dead time between exposures was $\sim 25$ ms.

All of these data were reduced using the ULTRACAM pipeline software. Debiassing, flatfielding and sky background subtraction were performed in the standard way. The source flux was determined with aperture photometry using a variable aperture, whereby the radius of the aperture is scaled according to the full width at half maximum (FWHM). Variations in observing conditions were accounted for by determining the flux relative to a comparison star in the field of view. We flux calibrated our targets by determining atmospheric extinction coefficients in each of the bands in which we observed and calculated the absolute flux of our targets using observations of standard stars (from \citealt{smith02}) taken in twilight. Using our extinction coefficients we extrapolated all fluxes to an airmass of $0$. 

\begin{figure*}
\begin{center}
 \includegraphics[width=0.99\textwidth]{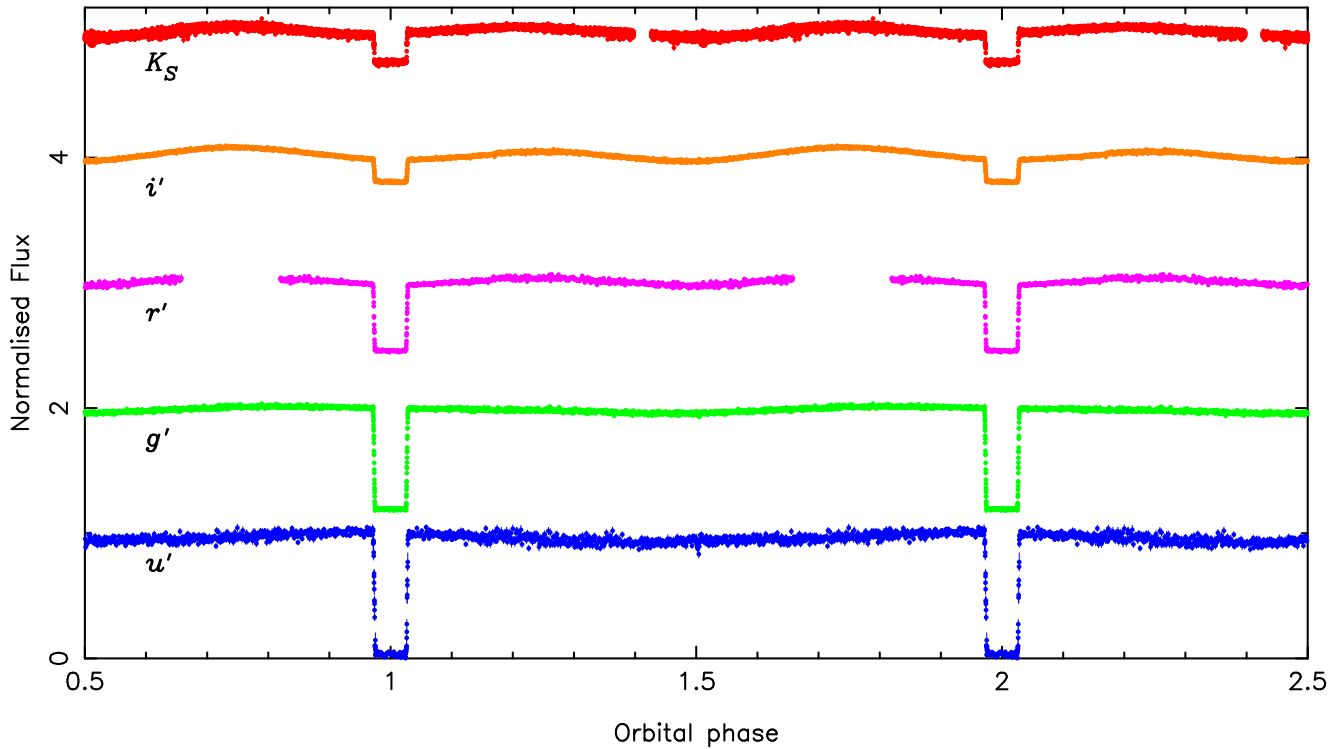}
 \caption{Phase folded ULTRACAM and HAWK-I light curves. The light curves have been normalised to the regions immediately before and after the primary eclipse and offset from each other vertically by 1. The gaps in the $r'$ and $K_\mathrm{S}$ bands are caused by a lack of coverage.}
 \label{fig:lcurves}
 \end{center}
\end{figure*}

\subsection{HAWK-I NIR photometry}

We observed SDSS\,J0303+0054 for a full binary orbit with the infrared imager HAWK-I installed at the Nasmyth focus of VLT-UT4 at Paranal \citep{kissler08}. We used the fast photometry mode which allowed us to window the detectors and achieve a negligible dead time between frames (a few microseconds). We used the $K_\mathrm{S}$ band filter and exposure times of 1 second. The data were reduced using the ULTRACAM pipeline in the same manner as described in the previous section. The data were flux calibrated using 2MASS measurements of nearby comparison stars. 

\subsection{X-shooter spectroscopy}

SDSS\,J0303+0054 was observed for almost 2 full binary orbits using the medium resolution spectrograph X-shooter \citep{dodorico06} mounted at the Cassegrain focus of VLT-UT2 at Paranal. X-shooter consists of 3 independent arms that give simultaneous spectra longward of the atmospheric cutoff (0.3 microns) in the UV (the ``UVB'' arm), optical (the ``VIS'' arm) and up to 2.5 microns in the near-infrared (the ``NIR''arm). We used slit widths of 1.0'', 0.9'' and 0.9'' in X-shooter's three arms and binned by a factor of two in the dispersion direction in the UVB and VIS arms resulting in a spectral resolution of 2500--3500 across the entire spectral range.

The reduction of the raw frames was conducted using the standard pipeline release of the X-shooter Common Pipeline Library (CPL) recipes (version 1.5.0) within ESORex, the ESO Recipe Execution Tool, version 3.9.6. The standard recipes were used to optimally extract and wavelength calibrate each spectrum. The instrumental response was removed by observing the spectrophotometric standard star LTT\,3218 and dividing it by a flux table of the same star to produce the response function. The wavelength scale was also heliocentrically corrected. A telluric correction was applied using observations of the DC white dwarf GD 248 obtained just before the start of our observations of SDSS\,J0303+0054.  

\section{Results}

\subsection{Light curves} \label{sec:lcurves}

\begin{figure}
\begin{center}
 \includegraphics[width=0.99\columnwidth]{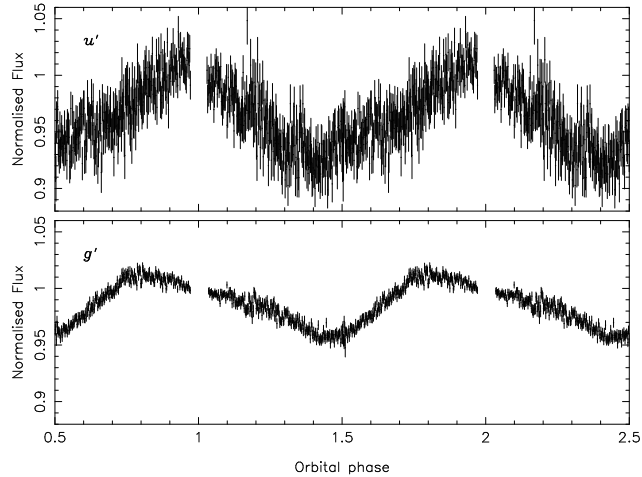}
 \caption{Out-of-eclipse variations in the ULTRACAM $u'$ and $g'$ band light curves. Eclipse points have been removed and the data binned by a factor of 5. There is a negligible contribution from the M star in the $u'$ band, hence the variations in this band must originate from the white dwarf. There is a small contribution from the M star in the $g'$ band in the form of ellipsoidal modulation, which increases the flux around phase 0.25 and 0.75, moving the maximum earlier and the minimum later, compared to the $u'$ band.}
 \label{fig:lc_zoom}
 \end{center}
\end{figure}

Figure~\ref{fig:lcurves} shows our ULTRACAM and HAWK-I light curves folded on the ephemeris of \citet{pyrzas09}. The features visible are similar to those reported by \citet{parsons10}, namely ellipsoidal modulation in the longer wavelength bands, and small out-of-eclipse variations in the $u'$ and $g'$ bands. Figure~\ref{fig:lc_zoom} shows a zoom-in on these variations, which are roughly sinusoidal, peaking just before the eclipse of the white dwarf. This variation is strong in the $u'$ band ($\sim$10\%) whilst it is slightly weaker in the $g'$ band ($\sim$5\%).

The M dwarf is not detected during the $u'$ band eclipse implying that it contributes a negligible amount of flux in this band. Therefore, the variations in the $u'$ band light curve must originate from variations in brightness on the surface of the white dwarf (ruling out starspots on the M star as suggested in \citealt{parsons10}). The fact that this variation has remained constant over the 5 years spanning our ULTRACAM observations strongly implies that the white dwarf is synchronously rotating with the binary (as is the M star, since its tidal synchronisation timescale is much shorter than the cooling age of the white dwarf). The M star contributes 20\% of the overall flux in the $g'$ band meaning that, in addition to the variation from the white dwarf, there is also a small ellipsoidal modulation component in this band (although the white dwarf variations still dominate). Since ellipsoidal modulation causes peaks at phases 0.25 and 0.75, the maximum of the light curve occurs slightly earlier than in the $u'$ band, and the minimum occurs slightly later (see Figure~\ref{fig:lc_zoom}).

\begin{figure*}
\begin{center}
 \includegraphics[width=0.99\textwidth]{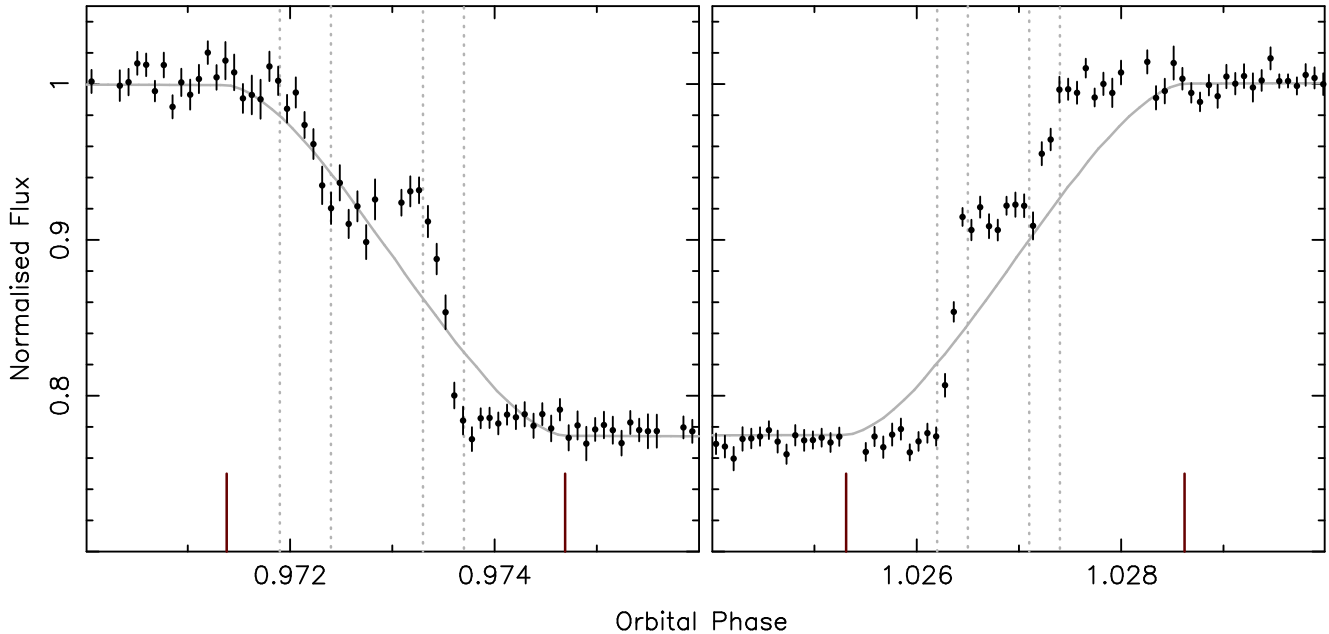}
 \caption{HAWK-I $K_\mathrm{S}$ band light curve around the time of the white dwarf eclipse. Overplotted is a model (gray line) of the white dwarf eclipse from fitting the $g'$ band light curve (i.e. the white dwarf photosphere) scaled to give the same eclipse depth. The vertical dotted lines indicate the contact phases of the eclipse of each pole, note the difference between these and the eclipse of the photosphere (the measured photospheric contact points are indicated by the red lines). Furthermore, there is a hiatus halfway through the ingress and egress, indicating that what we see eclipsed in the data are two small regions on the surface of the white dwarf.}
 \label{fig:hawk_ecl}
 \end{center}
\end{figure*}

Figure~\ref{fig:hawk_ecl} shows the HAWK-I $K_\mathrm{S}$ band light curve around the time of the white dwarf eclipse. The first thing to note is that the eclipse of a $\sim$8000\,K white dwarf photosphere \citep{pyrzas09} by an M4 dwarf should be $\sim$1\% in the $K_\mathrm{S}$ band (see for example Figure~\ref{fig:wdspec}), instead we observed a $\sim$20\% deep eclipse. Also plotted in Figure~\ref{fig:hawk_ecl} is a model fitted to the $g'$ band ULTRACAM light curve (see Section~\ref{sec:lc_model} for details of the model fitting). The contact phases of the white dwarf eclipse at optical wavelengths do not correspond to those observed in the $K_\mathrm{S}$ band. Furthermore, the ingress and egress are stepped, indicating that two small regions (bright in the $K_\mathrm{S}$ band) on the surface of the white dwarf are being eclipsed.

Taken together, our light curves strongly imply that the white dwarf in SDSS\,J0303+0054 is in fact a magnetic white dwarf accreting a small amount of material from the wind of the M star. In the $K_\mathrm{S}$ band we see cyclotron emission from accretion column above the magnetic poles of the white dwarf, it is this emission that is eclipsed. At optical wavelengths we see photospheric emission from the two heated pole caps on the white dwarf, on top of the general photospheric emission of the white dwarf (which dominates). Differences in the viewing angle to these poles as the white dwarf rotates causes the variations seen in the $u'$ and $g'$ band light curves. Although since the white dwarf photosphere dominates at these wavelengths, there is no hiatus visible during the eclipse at these wavelengths. Heated pole caps have been detected from every accreting magnetic white dwarf with appropriate observations (see for example \citealt{stockman94,araujo05,gansicke06,burleigh06}), hence there is little doubt that the white dwarf in SDSS\,J0303+0054 is accreting material from the wind of its companion.

A magnetic white dwarf also explains the infrared excess identified initially by \citet{debes12}. This excess is consistent with cyclotron emission from low-level accretion on to the magnetic white dwarf, and thus there is no compelling evidence for the existence of a circumbinary disk.

\subsection{Location of the poles}

The correct determination of the pole locations require the ephemeris of the HAWK-I data to be as accurate as possible, since a shift of only a couple of seconds is significant compared to the ingress/egress times. Fortunately we have ULTRACAM observations from a month before the HAWK-I observations, which reduces the uncertainty in the ephemeris for these data. We used the quadratic ephemeris derived in Section~\ref{sec:period} to determine the orbital phase of each HAWK-I data point. We estimate the timing accuracy to be better than 1 second, smaller than our HAWK-I exposure times, and hence not a large source of uncertainty.

\begin{figure}
\begin{center}
 \includegraphics[width=0.94\columnwidth]{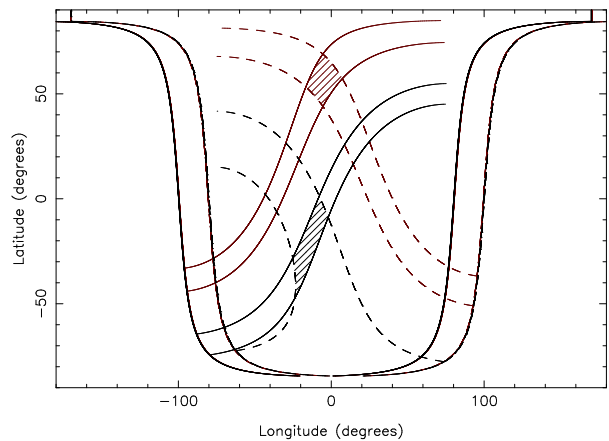}
 \caption{Location of the two magnetic poles on the surface of the white dwarf for an inclination of 84.5$^\circ$, longitude zero points towards the M star. Dashed lines show the position of the eclipse terminator during the ingress, as it moves north-eastward, whilst solid lines show the terminator during the egress, as it moves south-eastward. We show the terminator position for four phases on the ingress and four on the egress. These are the start and end phases of the eclipse of each pole as determined from the HAWK-I eclipse (the dotted lines in Figure~\ref{fig:hawk_ecl}). The shaded regions indicate the possible locations of the spots, black lines for the first (weaker) pole eclipsed (and last to emerge) and red lines for the second (stronger) pole to be eclipsed (the first to emerge). Also shown is the total surface of the white dwarf visible at these phases, indicated by the lines on either side, the difference between results from the rotation of the white dwarf during the eclipse.}
 \label{fig:spotmap}
 \end{center}
\end{figure}

We constrain the location of the poles on the surface of the white dwarf using the HAWK-I eclipse light curve and our model fit to the optical data (Section~\ref{sec:lc_model}). We use our model to determine the location of the eclipse terminator on the surface of the white dwarf at the start and end of the eclipse of each pole. Figure~\ref{fig:spotmap} shows the regions where the poles are located. It is a highly non-dipolar field with each pole pointing more-or-less in the direction of the M star. The first pole to be eclipsed is the weaker, southern pole (which gives a shallower eclipse depth). However, this is the last pole to emerge from behind the M dwarf. This places an upper limit on the binary inclination (87.3$^\circ$), above which it is impossible for this spot to emerge last, and the location of both spots is undetermined.

\subsection{Spectrum}

\begin{figure*}
\begin{center}
 \includegraphics[width=0.95\textwidth]{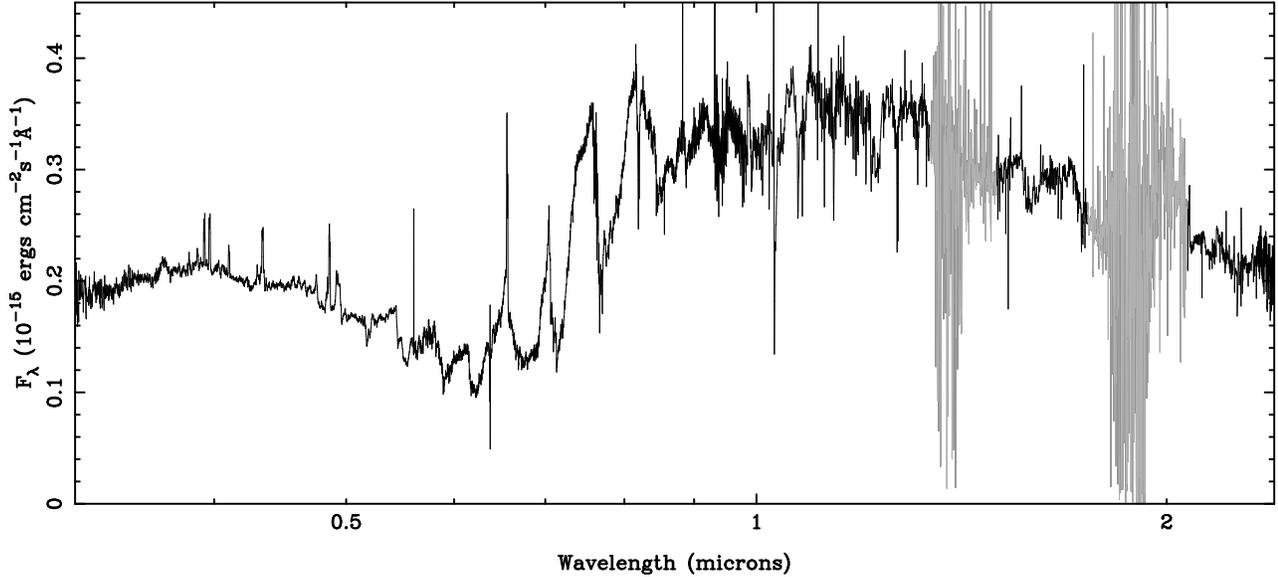}
 \caption{Averaged, telluric corrected X-shooter spectrum of SDSS\,J0303+0054. The strong $J$ and $H$ band telluric absorption regions have been grayed out.}
 \label{fig:avspec}
 \end{center}
\end{figure*}

\begin{figure}
\begin{center}
 \includegraphics[width=\columnwidth]{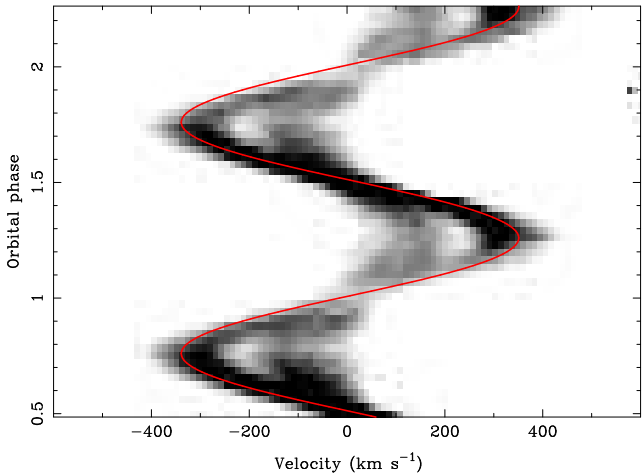}
 \caption{Trailed spectrum of the H$\alpha$ line. The continuum has been subtracted. Two emission components are visible, the strongest, high-amplitude component originates from the M dwarf and becomes stronger around phase 0.5. The weaker low-amplitude component originates between the two stars and is likely caused by magnetically confined material. The red line tracks the motion of the centre-of-mass of the M star.}
 \label{fig:halpha}
 \end{center}
\end{figure}

The average spectrum of SDSS\,J0303+0054 is shown in Figure~\ref{fig:avspec}. The two components are easily visible, although no absorption features from the white dwarf are detected, confirming its DC nature. Several strong emission lines are seen originating from the M dwarf including the hydrogen Balmer series and the \Ion{Ca}{ii} H and K lines. Figure~\ref{fig:halpha} shows a trailed spectrum of the H$\alpha$ emission line, revealing that it is in fact split in two obvious components. The strongest component comes from the M dwarf and peaks around phase 0.5, its radial velocity amplitude is slightly less than that of the M dwarf centre-of-mass (see Section~\ref{sec:rvs}). This implies that a large amount of the line flux originates from the inner hemisphere of the M star. This is often seen is systems with hot white dwarfs (e.g. \citealt{parsons10nn}), but the white dwarf in SDSS\,J0303+0054 is too cool to directly heat the M dwarf to such an extent (it receives $\sim$0.001 times its own luminosity from the white dwarf). This effect has been seen in other cool white dwarf plus main-sequence binaries, WD\,0137-349 for example \citep{maxted06}. It is possible that the accretion of the M dwarf's wind material on to the white dwarf produces X-rays directed back towards the M star, driving this emission on the hemisphere facing the white dwarf.

Arguably the most interesting feature in Figure~\ref{fig:halpha} is the second, weaker and lower-amplitude component. Given that the stronger component comes from the inner face of the M star, this weaker component cannot come from the M dwarf itself, but rather material between the two stars. However, since this weaker component moves in the same direction as the M dwarf, the material does not extend beyond the binary centre-of-mass towards the white dwarf. If it did we would expect to see some material moving in antiphase with the M star, which we do not detect. Therefore, it is likely that this material is being magnetically confined between the two stars. Similar features have been seen in several CV and have been attributed to slingshot prominences \citep{steeghs96,gansicke98,kafka05,kafka06,watson07}, similar features have also been seen in the PCEB QS\,Vir \citep{parsons11}. Interestingly, this weaker component shows a narrow eclipse-like feature around phase 0.79. Given that this feature repeats itself between orbits it is likely real. However, since the two stars are side-by-side at this phase, as seen from Earth, any eclipse of this material by one of the stars seems unlikely. The emission also appears to have some structure but the resolution of our data is inadequate to resolve this.

\subsection{M star radial velocity amplitude} \label{sec:rvs}

\begin{figure}
\begin{center}
 \includegraphics[width=0.99\columnwidth]{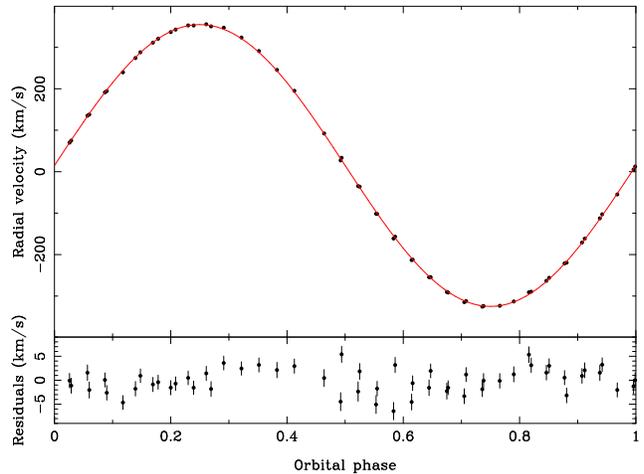}
 \caption{Radial velocity fit to the \Ion{Na}{i} 8200\,{\AA} absorption doublet from the M dwarf. The error bars are too small to be seen in the upper panel.}
 \label{fig:sodium}
 \end{center}
\end{figure}

We measure the radial velocity amplitude of the M dwarf using the \Ion{Na}{i} 8200\,{\AA} absorption doublet. We fitted the two lines separately, with a combination of a first order polynomial and a Gaussian component. We also fitted the \Ion{K}{i} 7700\,{\AA} absorption line (which is unaffected by telluric absorption) in the same manner. All the features gave consistent results and Figure~\ref{fig:sodium} shows the fit to one of the sodium lines. We measure a radial velocity amplitude for the M star of $K_\mathrm{sec}=339.9\pm0.3$\,{\kms} with a systemic component of $\gamma_\mathrm{sec}=14.9\pm0.2$\,\kms, consistent with, but more precise than, the measurements of \citet{pyrzas09}. We also find no significant  variations in the equivalent width of the line throughout the orbit, indicating that the strength of the absorption is uniform across the surface of the M star and hence we are indeed tracking the centre of mass of the star. 

\subsection{The white dwarf's spectrum}

Having determined the radial velocity amplitude of the M dwarf we then attempted to remove this component from the spectra to reveal the underlying white dwarf's spectrum. Two spectra were taken completely in eclipse and only contain the spectrum of the M dwarf, hence we averaged these spectra to make the M dwarf spectrum. We then shifted and subtracted this spectrum from all the others, based on our radial velocity measurements. In each case the M dwarf spectrum was scaled in order to minimise the residuals after subtraction in regions dominated by large M dwarf absorption features (regions also unlikely to contain any white dwarf features). This corrects for the varying contribution of the M dwarf component to the overall flux during the orbit, due mainly to ellipsoidal modulation. However, the M dwarf emission lines are not removed since their strengths vary during the orbit in a different way to the the absorption features. Furthermore, since the additional emission component from the confined material does not move with the M dwarf it too was not removed. To avoid confusing residuals we removed the emission components in the M dwarf template spectrum by manually setting the emission regions to the surrounding continuum level.

\begin{figure*}
\begin{center}
 \includegraphics[width=0.89\textwidth]{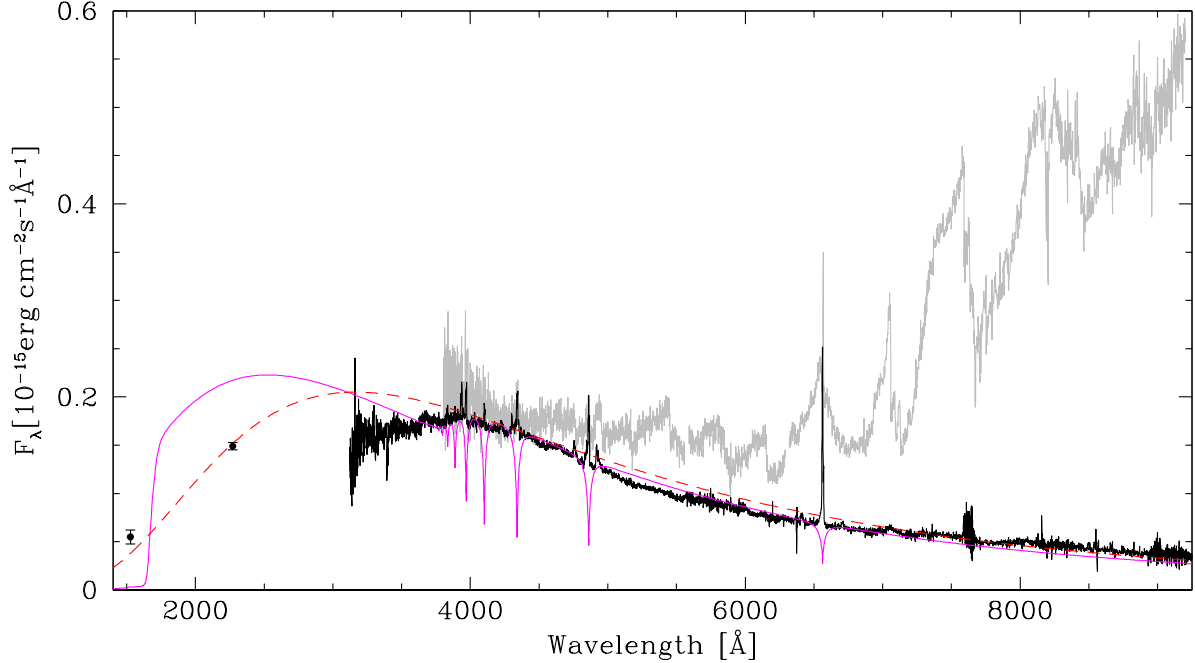}
 \caption{Spectrum of SDSS\,J0303+0054 with the white dwarf and M star components separated. Emission features from the M star and the confined material were not removed from the white dwarf component. Additional emission features are seen near the hydrogen Balmer lines (most obvious for the H$\beta$ line in this plot) these are Zeeman split and shifted components from the white dwarf. Also overplotted are a 9150\,K blackbody spectrum (dashed red line) and a non-magnetic DA white dwarf spectrum (solid magenta line) scaled to fit the spectrum of the white dwarf in SDSS\,J0303+0054. We also show the GALEX FUV and NUV fluxes, the non-magnetic white dwarf clearly over-predicts the NUV flux.}
 \label{fig:wdspec}
 \end{center}
\end{figure*}

Figure~\ref{fig:wdspec} shows the white dwarf's spectrum in the UVB and VIS arms of X-shooter. SDSS\,J0303+0054 was observed with GALEX as part of the Medium Imaging Survey (MIS) with exposure times of 2422s and 3687s in the far- and near-ultraviolet, respectively. We indicate the measured GALEX fluxes in  Figure~\ref{fig:wdspec}. As previously noted, the strong M dwarf emission components have not been removed, neither has the additional component from the confined material. However, additional emission lines are seen in Figure~\ref{fig:wdspec}. These features do not move with the M dwarf. Given the magnetic nature of the white dwarf, these are likely to be Zeemen split and shifted components of the hydrogen lines from the white dwarf. Zeeman split emission lines have only been seen in one other white dwarf, GD\,356 \citep{greenstein85,ferrario97}. The emission lines in the apparently isolated white dwarf GD\,356 are the result of a temperature inversion in its atmosphere, though the origin of this inversion remains unclear. \citet{wickramasinghe10} placed an upper limit for the mass of any companion to GD\,356 of 12 Jupiter masses and suggested that accretion was unlikely to be the source of the emission lines. However, since we know there is some low-level accretion on to the white dwarf in SDSS\,J0303+0054 via the wind of the M star, it is likely that this accreted material creates a shock at or near the surface of the white dwarf. This hot shock material cools and sinks towards the centre of the white dwarf creating a temperature inversion, which is responsible for the emission lines. It is likely that these emission features mask the underlying absorption components, as was revealed in the polarisation spectra of GD\,356 \citep{ferrario97}. Unfortunately these emission features are too weak to give a reliable estimate of the radial velocity amplitude of the white dwarf. Furthermore, it is likely that variations in the magnetic field strength (hence Zeeman effect) across the surface of the white dwarf would make these components unreliable tracers of the centre-of-mass of the white dwarf.

\begin{figure*}
\begin{center}
 \includegraphics[width=0.99\textwidth]{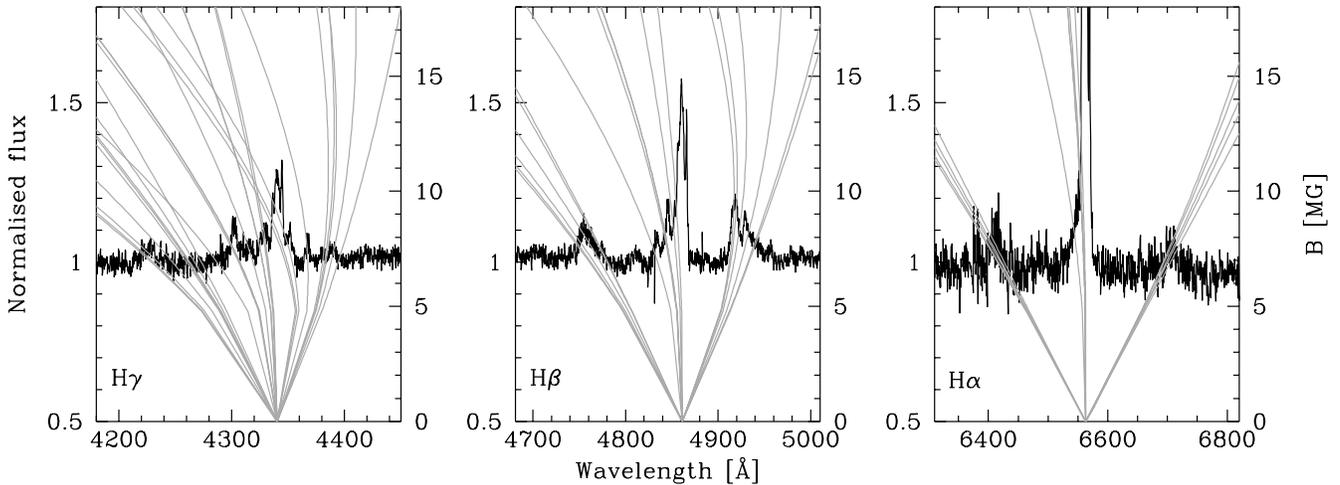}
 \caption{White dwarf emission features. The central (strong) components in each line are from the M star and confined material. Overplotted are the positions of the Zeeman split components as a function of the magnetic field strength. The peaks of the emission features show good agreement with a field strength of $8\pm1$\,MG.}
 \label{fig:wdemis}
 \end{center}
\end{figure*}

Figure~\ref{fig:wdemis} shows a zoom in on several of the hydrogen Balmer lines. Over-plotted are the locations of the Zeeman split components as a function of magnetic field strength from \citet{friedrich96}. Good agreement is found for a field strength of 8\,MG. This field strength is supported by the lack of any obvious cyclotron lines in our data. For a field strength of 8\,MG the fundamental cyclotron line will be located at 13.3\,microns far beyond our wavelength coverage, in fact the 5$^\mathrm{th}$ harmonic at 2.2\,microns would be the lowest harmonic visible. This is consistent with the large infrared excess detected by \citet{debes12} which peaks around 10\,microns.

The emission lines from the white dwarf likely originate from above its surface. Although measurements of the gravitational redshift of absorption and emission lines in the PCEB LTT\,560 showed little difference \citep{tappert11} indicating that the emitting regions are close to the surface. Since the magnetic field strength decreases with hight as $1/r^3$, 8\,MG is likely a lower limit on the field strength. However, the components are fairly sharp meaning that the field strength does not vary considerably over the emitting region, hence it cannot extend too far. Furthermore, in polars, where both (photospheric) Zeeman components and cyclotron lines are visible, the field strengths generally agree. For example, a field strength of 24--25\,MG was measured from both the Zeeman and cyclotron components in the polar MR Ser \citep{schwope93} and the 13\,MG measurement of the field strength in EF Eri \citep{beuermann07} from the Zeeman lines is consistent with the 12.6\,MG measurement from the cyclotron lines \citep{campbell08}. Therefore, it is likely that the field strength of the white dwarf in SDSS\,J0303+0054 is close to 8\,MG.

In Figure~\ref{fig:wdspec} we also plot a blackbody spectrum and a non-magnetic DA white dwarf spectrum fitted to the observed white dwarf spectrum. Adopting a distance of 140\,pc and a white dwarf radius of 0.00975\,\RSUN (see section \ref{sec:sys_paras}) gives a best fit blackbody temperature of 9150\,K. A DA white dwarf model with a $T_\mathrm{eff}$ of 9150\,K (shown in Figure~\ref{fig:wdspec}) matches the optical spectrum, but clearly over-predicts the GALEX NUV flux. This is somewhat unexpected as the effect of a 8\,MG field should not dramatically affect the Lyman $\alpha$ line. However, with the data to hand we are unable to explain this discrepancy. UV spectroscopy would be helpful in identifying the cause.

\subsection{M star rotational broadening}

\begin{figure}
\begin{center}
 \includegraphics[width=0.99\columnwidth]{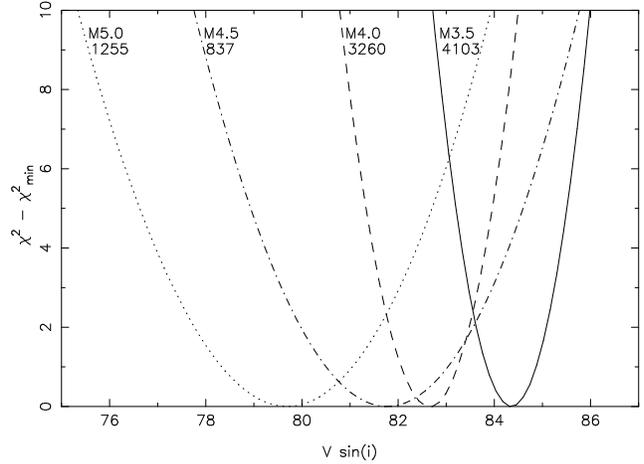}
 \caption{$\chi^2$ as a function of rotational broadening for the M star in SDSS\,J0303+0054, using four different template stars, using a limb-darkening coefficient of 0.55. The number shown below the spectral type is the minimum $\chi^2$ value for that template. The M4.5 template broadened by 81.7\,{\kms} gives an excellent fit to the M dwarf in SDSS\,J0303+0054.}
 \label{fig:rbroad}
 \end{center}
\end{figure}

Limits can be placed on the physical parameters of the two stars in SDSS\,J0303+0054 by measuring the rotational broadening ($v_\mathrm{rot}\sin{i}$) of the M star. This was achieved using the method detailed in \citet{marsh94}. We corrected the radial velocity of each spectrum to the rest frame of the M star and averaged the resulting spectra together. We then artificially broadened the spectra of four template M dwarf stars: GJ 849 (M3.5), GJ 876 (M4.0), GJ 3366 (M4.5) and GJ 2045 (M5.0), and subtracted these from the spectrum of SDSS\,J0303+0054 (multiplied by a constant representing the fraction of the flux from the M star). The constant was varied to optimise the subtraction, and the $\chi^2$ difference between the residual spectrum and a smoothed version of itself is computed. We focused on the sodium 8190{\AA} absorption doublet (using the wavelength range 8100--8300{\AA}). Figure~\ref{fig:rbroad} shows the $\chi^2$ of the fit as a function of rotational broadening for all four template stars as well as the minimum $\chi^2$ values. We found that the M3.5 and M4.0 template spectra were a poor fit to the M star in SDSS\,J0303+0054, giving a minimum $\chi^2$ a factor of five and four worse than the M4.5 template respectively. The M5.0 template was a better fit than both these templates but not as good as the M4.5 template (by a factor of 1.5 in $\chi^2$) which gives an excellent fit to the data, although the optimal values for all the templates are similar. These values are slightly dependent upon the choice of limb-darkening coefficient used. For a main-sequence star with a spectral type in this range we would expect a limb-darkening coefficient of $\sim$0.5 \citep{claret11} around the sodium lines. We note that this value is strictly only appropriate for the continuum of the M star and so we calculated the optimal values for a range of limb-darkening coefficients from 0.4 to 0.6 and find that this shifts the results by $\sim$1\kms, similar to the small effects seen by \citet{copperwheat12} for the donor star in the CV OY\,Car. Therefore, we conclude that the M star in SDSS\,J0303+0054 has a spectral type of M4.5--M5.0 with a rotational broadening of $v_\mathrm{rot}\sin{i}=81.7\pm1.1$\,{\kms} (the uncertainty comes from the unknown limb-darkening coefficient, the formal error estimate is $\Delta\chi^2\pm1=0.7$ for the M4.5 template star). We will use this measurement to constrain the parameters of the stars in Section~\ref{sec:sys_paras}.

\subsection{Modeling the light curve} \label{sec:lc_model}

Eclipsing PCEBs have been used to measure masses and radii of white dwarfs and low-mass stars to high-precision (see for example \citealt{parsons10nn,parsons12gk,parsons12,pyrzas12}). For SDSS\,J0303+0054, we chose to fit just the $g'$ band light curve around the eclipse of the white dwarf. In theory the ellipsoidal modulation seen in the $i'$ band light curve can be used to constrain the inclination (and hence radii) of the system. However, the magnetic effects of the white dwarf on the light curves (Section~\ref{sec:lcurves}) are difficult to model correctly and could lead to significant systematic uncertainties in any derived parameters. The transit of the white dwarf across the face of the M star can also be used to measure the inclination but is too shallow to detect in our data. The eclipse of the white dwarf in the $g'$ band is deep and has high signal-to-noise, whilst also being less affected by the photospheric emission from the heated pole caps than the $u'$ band, making it the cleanest feature visible in our light curves. 

We model the light curve data using using a code written for the general case of binaries containing white dwarfs (see \citealt{copperwheat10} for a detailed description). The program subdivides each star into small elements with a geometry fixed by its radius as measured along the direction of centres towards the other star with allowances for Roche geometry distortion. The limb darkening of both stars was set using a 4-coefficient formula: 

\begin{equation}
I(\mu)/I(1) = 1-\sum\limits_{i=1}^4 a_i(1-\mu^{i/2}),
\end{equation}
where $\mu$ is the cosine of the angle between the line of sight and the surface normal. For the M star, we use the coefficients for a $T_\mathrm{eff}=2900\,K$, $\log{g}=5$ main sequence star \citep{claret11}. We use the coefficients for a $T_\mathrm{eff}=8000\,K$, $\log{g}=8$ white dwarf from \citet{gianninas13}. We initially fitted each eclipse individually using Levenberg-Marquardt minimisation in order to measure precise mid-eclipse times and hence accurately phase-fold the data. The measured mid-eclipse times, as well as other published eclipse times for SDSS\,J0303+0054 are listed in Table~\ref{tab:etimes}.  

\begin{table}
 \centering
  \caption{Eclipse times for SDSS\,J0303+0054. (1) \citet{pyrzas09}, (2) \citet{parsons10}, (3), this paper, (4) \citet{backhaus12}.}
  \label{tab:etimes}
  \begin{tabular}{@{}lcc@{}}
  \hline
Cycle & MJD(BTDB)         & Reference \\ 
No.   & (mid-eclipse)     &           \\
 \hline
0     & 53991.11741(20)   & (1)       \\
14    & 53992.99923(20)   & (1)       \\
23    & 53994.20929(20)   & (1)       \\
44    & 53997.03229(20)   & (1)       \\
2559  & 54335.14302(20)   & (1)       \\
2589  & 54339.17583(20)   & (1)       \\
2960  & 54389.05324(20)   & (1)       \\
2968  & 54390.1282934(20) & (2)       \\
2976  & 54391.2037900(18) & (2)       \\
3058  & 54402.2276842(228)& (2)       \\
11300 & 55510.2629766(22) & (3)       \\
11307 & 55511.2040401(22) & (3)       \\
11411 & 55525.1855628(40) & (3)       \\
13443 & 55798.362876(13)  & (4)       \\
13510 & 55807.370189(14)  & (4)       \\
13533 & 55810.462273(12)  & (4)       \\
13874 & 55856.305526(11)  & (4)       \\
13897 & 55859.397585(11)  & (4)       \\
13926 & 55863.296278(10)  & (4)       \\
13948 & 55866.253894(13)  & (4)       \\
16283 & 56180.1658250(19) & (3)       \\
16505 & 56210.0109853(24) & (3)       \\
16535 & 56214.0441160(15) & (3)       \\
\hline
\end{tabular}
\end{table}

\begin{figure}
\begin{center}
 \includegraphics[width=0.99\columnwidth]{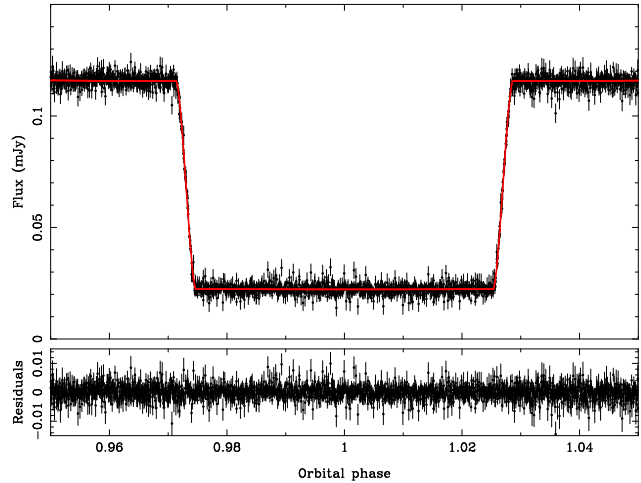}
 \caption{Model fit to the eclipse of the white dwarf in the ULTRACAM $g'$ band.}
 \label{fig:ecl_fit}
 \end{center}
\end{figure}

Having determined the mid-eclipse times for all our light curves we phase-folded the data and fitted a model to the combined light curve. The eclipse of the white dwarf alone does not contain enough information to break the degeneracy between the orbital inclination($i$) and the radii of the two stars scaled by the orbital separation ($R_\mathrm{WD}/a$ and $R_\mathrm{sec}/a$). However, for a given inclination and mass ratio, the eclipse light curve (combined with the period and $K_\mathrm{sec}$ measurement) allows us to solve for the masses and radii of both stars using only Kepler's laws. Therefore, we fitted the eclipse light curve over a range of probable inclinations ($i>80^\circ$) and mass ratios ($q=M_\mathrm{sec}/M_\mathrm{WD}<0.35$). Figure~\ref{fig:ecl_fit} shows the fit to the combined $g'$ band eclipse light curve. 

\section{Discussion} 

\subsection{System parameters} \label{sec:sys_paras}

\begin{figure}
\begin{center}
 \includegraphics[width=0.99\columnwidth]{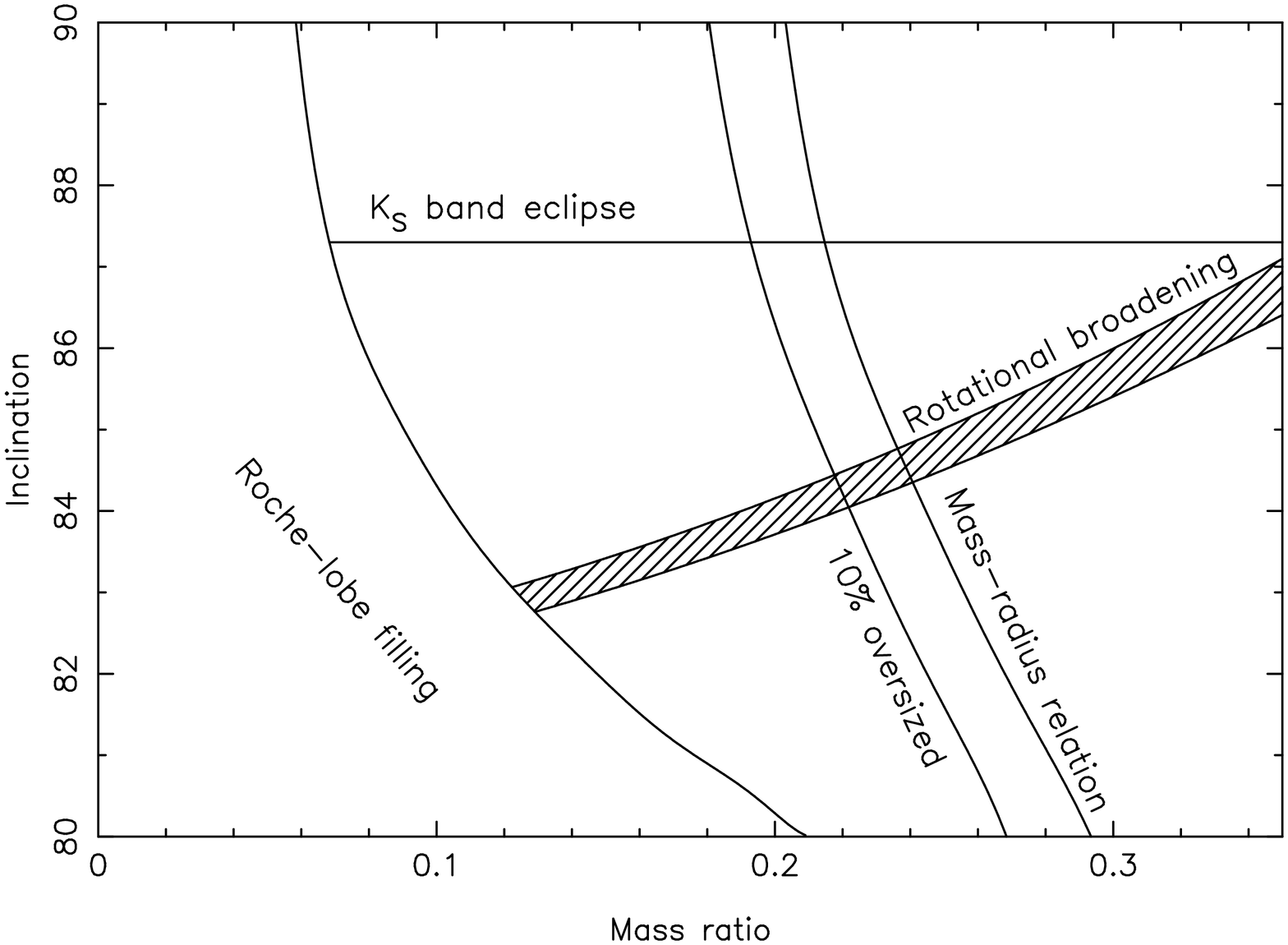}
 \caption{Constraints on the inclination and mass ratio of the binary. The light curve model constrains the minimum mass ratio, below which the M star fills its Roche lobe (i.e. the area to the right of this line is consistent with the $g'$ band eclipse data). The shaded area shows the region that is also consistent with the rotational broadening measurement. The $K_\mathrm{S}$ band eclipse places an upper limit on the inclination, above which it is impossible to determine the location of the poles on the surface of the white dwarf. This constraint is only loosely dependent upon the mass ratio. Also plotted is the mass radius relationship for a 3\,Gyr old active star star from \citet{morales10} and the same relationship but with a 10\% oversized main-sequence star.}
 \label{fig:sys_paras}
 \end{center}
\end{figure}

As previously stated, modeling the eclipse of the white dwarf alone does not allow us to independently solve for all the system parameters. However, it does allow us to exclude some areas of parameter space. This is because it is impossible to fit the eclipse light curve for low mass ratios without the M star filling its Roche-lobe, which we know not to be the case (there are no disc or stream features in our data and SDSS\,J0303+0054 is not an X-ray source). Figure~\ref{fig:sys_paras} shows the region in which the M star fills its Roche lobe, from this constraint alone we know that $q>0.06$. We also plot the upper limit on the inclination of $87.3^\circ$ from the $K_\mathrm{S}$ band eclipse of the two poles on the surface of the white dwarf, above which it is impossible to reproduce their observed eclipse phases in the $K_\mathrm{S}$ band eclipse light curve (i.e. their location on the surface of the white dwarf is undetermined).

For each combination of mass ratio and inclination we compute the volume average radius of the M star. We then determined what the rotational broadening of such a synchronously rotating M dwarf would be. The shaded area of Figure~\ref{fig:sys_paras} shows the region where this computed value is consistent with the measured value from our spectroscopic data (i.e. the true system parameters lie somewhere in the shaded region). This increases our minimum mass ratio limit to $q>0.12$ and places a lower limit on the inclination of $i \geq 83^\circ$. 

Also plotted in Figure~\ref{fig:sys_paras} is the mass-radius relationship for a 3\,Gyr old active low-mass star from \citet{morales10}. The masses and radii of low mass stars in eclipsing PCEBs have generally been consistent with these mass-radius relationships (see for example \citealt{parsons12}). However, several measurements from eclipsing PCEBs as well as results from eclipsing main-sequence binaries and other sources have shown that the measured radii can be up to 10\% oversized when compared to mass-radius relationships \citep{feiden12,terrien12}. Therefore, we also also show the constraints for a 10\% oversized M star in Figure~\ref{fig:sys_paras}, the actual M star radius almost certainly lies between these two limits. The region where these constraints overlap with the rotational broadening constraint indicates the most likely system parameters, which are listed in Table~\ref{tab:sys_paras}. These results are broadly consistent with those of \citet{pyrzas09} but are more precise. These parameters also predict an ellipsoidal modulation amplitude consistent with our ULTRACAM $i'$ band observations (where this effect dominates), although due to the heated poles of the white dwarf the fit is not perfect. The white dwarf's mass and radius show good agreement with models but our use of the low-mass star mass-radius relationship means that these results are not suitable for testing the relationships themselves. Nevertheless, there is no doubt that SDSS\,J0303+0054 hosts one of the most massive white dwarfs in a close, detached binary.

We estimate the distance to SDSS\,J0303+0054 following the prescription of \citet{beuermann06}. For M-dwarfs, the surface brightness near 7500\,\AA, and depth of the TiO band near 7165\,\AA\ are a strong function of the spectral type. \citet{beuermann06} provides a calibration of the surface brightness $F_\mathrm{TiO}$ defined as the the difference between the mean surface fluxes in the bands 7450-7550\,\AA\ and 7140-7190\,\AA. Measuring the observed flux $f_\mathrm{TiO}$ from the spectrum, the distance is then calculated as
\begin{equation}
d=\sqrt{R_\mathrm{sec}^2\frac{F_\mathrm{TiO}}{f_\mathrm{TiO}}}.
\end{equation}
The uncertainty on $R_\mathrm{sec}$ is fairly small hence the main uncertainties in the distance estimate are the flux calibration of the spectroscopy, and the spectral type of the companion. Adopting a conservative uncertainty in the spectral type of M$5\pm1.0$, and using the X-shooter spectrum, we find a distance of $140\pm25$\,pc, which is consistent with the result of \citet{debes12}. 

Better (model-independent) constraints on the system parameters can be obtained using the full light curve (i.e. the ellipsoidal modulation) but will require a better understanding of the variations arising from the magnetic white dwarf (e.g. via polametric observations). This would lead to the first model-independent mass and radius measurement for a magnetic white dwarf.

\subsection{Location of the confined material}

We can use our constraints on the system parameters to calculate the position of the confined material seen in the H$\alpha$ line (Figure~\ref{fig:halpha}). Assuming that this material co-rotates with the binary, its location is given by
\begin{eqnarray}
K_\mathrm{mat}/K_\mathrm{sec} = (1+q) \left( \frac{R}{a} \right) - q,
\end{eqnarray}
where $K_\mathrm{mat}$ and $K_\mathrm{sec}$ are the radial velocity semi-amplitudes of the extra material and M star respectively, $q$ is the mass ratio and $R/a$ is the distance of the material from the white dwarf scaled by the orbital separation ($a$).

We measure $K_\mathrm{mat}$ by fitting the H$\alpha$ line with a combination of a first order polynomial and two Gaussians. We follow the procedure outlined in \citet{parsons12gk}, whereby all of the spectra are fitted simultaneously and the position of the Gaussians are varied from spectrum-to-spectrum according to their orbital phase. This approach minimises any systematic effects caused by the two components crossing over. We also allowed the strength of the M dwarf component to vary with orbital phase according to
\begin{eqnarray}
H = H_0 - H_1\cos(2 \pi \phi), \nonumber
\end{eqnarray}
which allows the strength to peak at phase 0.5.

We find $K_\mathrm{mat}=117\pm1$\,\kms. Using this measurement and the values in Table~\ref{tab:sys_paras} gives $R/a=0.47$, indicating that the material is located roughly half way between the two stars, slightly closer to the white dwarf, but still some way from the binary centre-of-mass.

\begin{table}
\caption{Orbital and physical parameters of SDSS\,0303+0054.}
\label{tab:sys_paras}
\begin{tabular}{@{}ll@{}}
\hline
Parameter                         & Value                   \\
\hline
T0 (MJD(BTDB))                    & $53991.117\,307(2)$     \\
Orbital period                    & $0.134\,437\,666\,8(1)$\,days      \\
$\dot{P}$                         & $1.616(4)\times10^{-11}$\,s s$^{-1}$ \\
$K_\mathrm{sec}$                    & $339.9\pm0.3$\,\kms     \\
$v_\mathrm{rot,sec}\sin{i}$         & $81.7\pm1.1$\,\kms      \\
Orbital separation                & 1.106--1.125\,\RSUN     \\
Orbital inclination               & 84.2$^\circ$--84.7$^\circ$ \\
Mass ratio ($M_\mathrm{sec}/M_\mathrm{WD}$) & 0.22--0.24              \\
White dwarf mass                  & 0.825--0.853\,\MSUN     \\
White dwarf radius                & 0.00971--0.00978\,\RSUN \\
White dwarf field strength        & $8\pm1$\,MG             \\
M star mass                       & 0.181--0.205\,\MSUN     \\
M star radius (volume-averaged)   & 0.216--0.221\,\RSUN     \\
M star spectral type              & M4.5--M5.0              \\
Distance                          & $140\pm25$\,pc          \\
\hline                                                                  
\end{tabular}
\end{table}

\subsection{Orbital period variations} \label{sec:period}

\begin{figure}
\begin{center}
 \includegraphics[width=0.99\columnwidth]{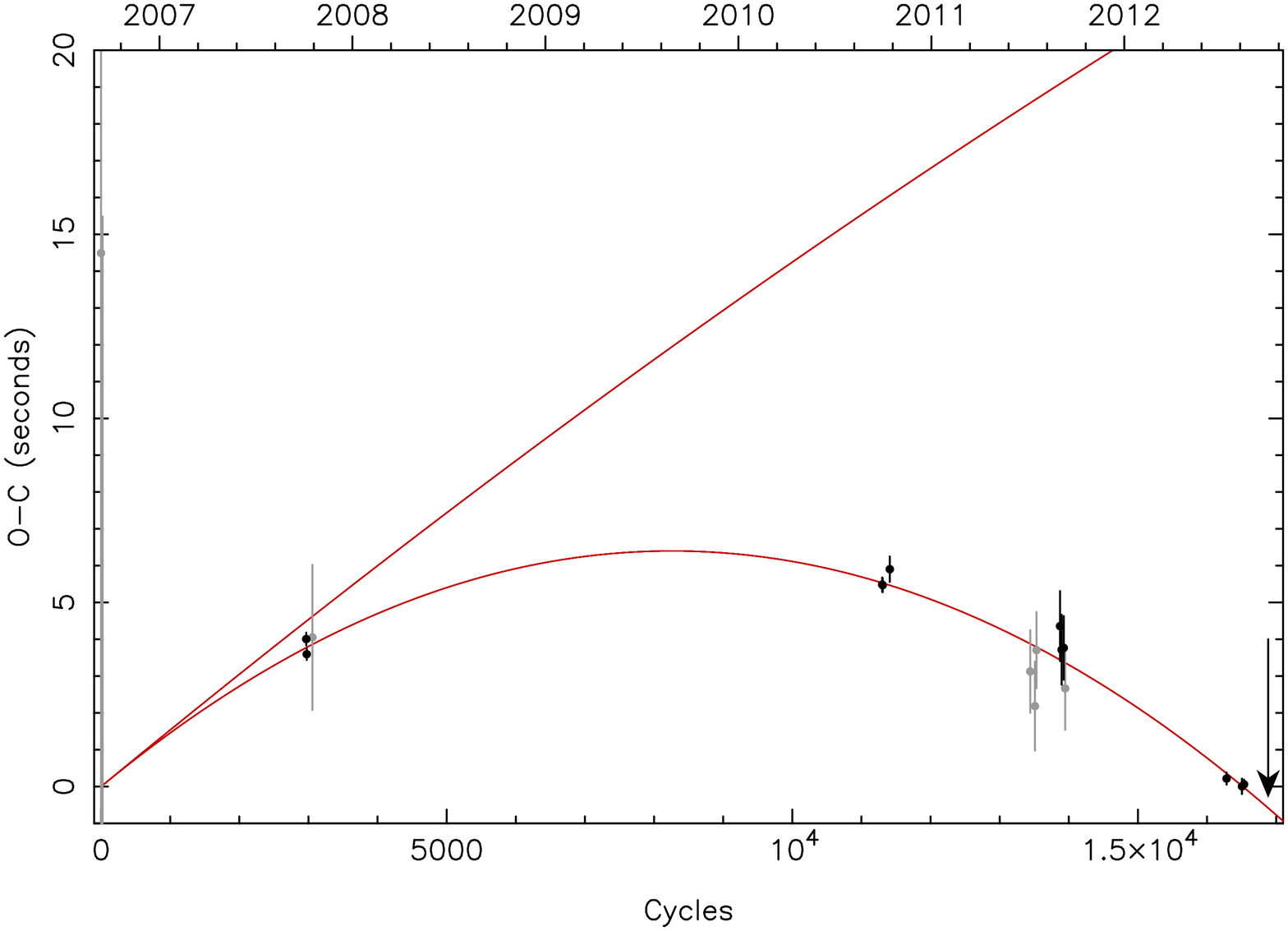}
 \caption{Observed-Calculated (O-C) plot for the eclipse arrival times of SDSS\,J0303+0054, with the gradient term removed. Times with uncertainties larger than 1 second have been grayed out. We also show a quadratic fit to the eclipse times. The other curve (virtually a straight line) is the maximum period change possible from magnetic braking alone, which vastly under-predicts the true change meaning an additional period change mechanism is required. The arrow indicates the time of the HAWK-I observations.}
 \label{fig:omc}
 \end{center}
\end{figure}

The best fit linear ephemeris to all the eclipse times of SDSS\,J0303+0054 is
\[\mathrm{MJD(BTDB)} = 53991.117\,307(2) + 0.134\,437\,666\,8(1)  E.\]
However, the eclipse arrival times show some deviations from this ephemeris ($\sim$5 seconds, or more than 30$\sigma$ for some points). Adding a quadratic term to the ephemeris results in a far better fit and is shown in Figure~\ref{fig:omc}. The best-fit quadratic ephemeris is
\begin{eqnarray}
\mathrm{MJD(BTDB)} & = & 53991.117\,243(3) + 0.134\,437\,687\,9(7) E \nonumber \\
                   &   & - 1.086(38) \times 10^{-12} E^2, \nonumber
\end{eqnarray}
which, following the procedure of \citet{brinkworth06}, implies a rate of period change of $\dot{P} = 1.616(4)\times10^{-11}$\,s s$^{-1}$ or an angular momentum loss of $\dot{J}/J \simeq 4.0 \times 10^{-11}$.

To estimate an upper limit on the angular momentum loss from the system, we use the standard magnetic braking relationship from \citet{rappaport83} (which is much larger than the angular momentum loss via gravitational radiation)
\begin{eqnarray}
\dot{J} \approx -3.8 \times 10^{-30} M_{\sun} R_{\sun}^{4} M_\mathrm{sec} R_\mathrm{sec}^{\gamma} {\omega}^{3} \, \mathrm{erg},
\end{eqnarray}
where $M_\mathrm{sec}$ and $R_\mathrm{sec}$ are the secondary star's mass and radius and $\omega$ is the angular frequency of rotation of the secondary star. $\gamma$ is a dimensionless parameter which can have a value between 0 and 4. We used the values from Table~\ref{tab:sys_paras} and $\gamma=0$ to maximise the angular momentum loss (since $R_\mathrm{sec}<1$) and plot the expected period variation in Figure~\ref{fig:omc}. The maximum angular momentum loss via standard magnetic braking in SDSS\,J0303+0054 is $\dot{J}/J \simeq 5.3 \times 10^{-12}$, an order of magnitude smaller than seen and the curve in Figure~\ref{fig:omc} clearly under-predicts the observed period change. 

The magnetic nature of the white dwarf may cause a reduction in angular momentum loss via magnetic braking. This is because the field lines from the M star are closed or connect to the field of the white dwarf, reducing the magnetic flux in open field lines hence reducing the effect of magnetic braking \citep{li94}. Furthermore, the mass of the M star is below the fully convective boundary hence it is likely that magnetic braking is not operating at all. Therefore, there is clearly an additional mechanism causing the period decrease. 

The standard mechanism used to explain period variations in close binaries is Applegate's mechanism \citep{applegate92}, which is caused by fluctuations in the gravitational quadrupolar moment of the M star. Using the approach of \citet{brinkworth06} we find that $1.2 \times 10^{36}$\,erg are required to drive the $\Delta P = 2.086 \times 10^{-4}$ second period variation over the last $\sim$6 years. Interestingly, the M star is comfortably able to supply this energy; the total energy available from the M star over this time period is $2.9 \times 10^{39}$\,erg, therefore Applegate's mechanism is a possible cause for the early eclipse arrival times. However, virtually every PCEB with long enough coverage eventually shows period variations that are too large to be explained as the result of Applegate's mechanism. In many cases the variations are interpreted as the influence of a third body in orbit around the binary (see \citealt{zorotovic13} for a comprehensive list). It remains to be seen what the true cause is of the period decrease in SDSS\,J0303+0054.

\subsection{Evolution}

We calculate the future evolution of SDSS\,J0303+0054 following \citet{schreiber03}. Given the low mass of the M star we assume that the only angular momentum loss mechanism that is acting is gravitational radiation. This allows us to estimate the time needed to evolve in to a CV to be slightly less than a Gyr ($\sim$970Myrs). Roche-lobe overflow will start when the orbital period has decreased to $P_\mathrm{orb}\sim2.1$\,hrs corresponding to a binary separation of 0.83\,\RSUN. At the onset of Roche-lobe overflow material will reach a minimum distance from the white dwarf of $r_\mathrm{min}/a = 0.0488 \times q-0.464$ \citep{lubow75}, which corresponds to roughly ten per cent of the binary separation of the CV that will have formed. Using Equation 1 of \citet{ferrario89} we estimate the equilibrium radius between magnetic pressure from the $\sim$8\,MG magnetic field of the white dwarf and the ram pressure from the infalling material to be smaller than $r_\mathrm{min}$ even for very low accretion rates of $10^{−14}$g/sec. This implies that a disk will form around the white dwarf and that SDSS\,J0303+0054 will become an intermediate polar. However, as a note of caution, we highlight that the lowest-field polar known, i.e. V2301 Oph \citep{ferrario95}, is quite similar to the CV that SDSS\,J0303+0054 will become. Furthermore, the equation from \citet{ferrario89} represents just a rough approximation as, for example, a pure dipole and radial inflow is assumed, and we know the field is not dipolar, based on the location of the poles. We therefore cannot completely exclude the possibility that SDSS\,J0303+0054 will become a polar instead of an intermediate polar.

At first glance, the history of SDSS\,J0303+0054 seems to be even less certain than its future. It could either have emerged from the common envelope and is now a classical PCEB evolving slowly into a CV or, given its current period of 3.2\,hrs and the period it will have when starting mass transfer of 2.1\,hrs, one might also speculate that it is in fact a detached CV that recently entered the period gap \citep{davis08}. However, one needs to take into account that above the orbital period gap SDSS\,J0303+0054 must have been an intermediate polar (i.e. the white dwarf was asynchronously rotating). According to Equation 1 of \citet{campbell84}, after entering the period gap it would have taken $\sim1.2$\,Gyrs for the white dwarf to reach its current synchronised state. If this is correct it implies that SDSS\,J0303+0054 is not a gap-CV: $1.2$\,Gyrs ago the system had an orbital period of $\sim$4.2\,hrs and was not even close to Roche-lobe filling, hence SDSS\,J0303+0054 has never been a CV. We therefore conclude that SDSS\,J0303+0054 is a genuine detached PCEB on its way from the common envelope phase towards becoming a CV. In this scenario, the cooling age of the WD ($\sim2$ Gyrs) tells us that roughly two thirds of its PCEB lifetime have passed (recall it needs $\sim1$Gyrs to start mass transfer). Given the high mass of the primary, the progenitor of the white dwarf must have been fairly massive ($>4$\,\MSUN). Thus, the time the system spent as a main sequence binary is small compared to the total PCEB lifetime ($\sim$3 Gyrs). This means that SDSS J0303+0054 is a pre-CV in the sense defined by \citet{schreiber03}, i.e. it's total CV formation time is less than the age of the Galaxy and it can be considered representative for the progenitors of the current CV population.

\section{Conclusions}

We have discovered that the white dwarf in the eclipsing post common envelope binary SDSS\,J0303+0054 is a magnetic white dwarf with a field strength of 8\,MG, measured from Zeeman emission components. This is only the second known white dwarf to show Zeeman split emission lines, after GD 356. The magnetic nature of the white dwarf naturally explains the infrared excess seen in the system without the need for a circumbinary disk. We also detected the eclipse by the M dwarf of two magnetic poles on the surface of the white dwarf. Using a model fitted to our optical data we were able to determine the location of these poles and found that the field is highly non-dipolar, with both poles facing more-or-less towards the M star. Our spectroscopic observations revealed the existence of material roughly half way between the M star and the white dwarf, that is likely to be magnetically confined.

We combine radial velocity measurements, rotational broadening measurements and a model to the light curve to constrain the system parameters. The 0.84\MSUN white dwarf is relatively massive for a post common envelope binary whilst the M4.5 dwarf star has a mass of 0.19\MSUN.

Finally, we also detect a decrease in the orbital period of the system over the last 6 years of $\Delta P = 2.086 \times 10^{-4}$\,s. The magnitude of this period decrease is too large to be caused by angular momentum loss via magnetic braking. However, it can be comfortably explained as the result of Applegate's mechanism. Further monitoring of the system will reveal the true nature of this period change. SDSS\,J0303+0054 is a pre-CV that will likely evolve into an intermediate polar in $\sim$970\,Myrs.

\section*{Acknowledgments}

We thank the anonymous referee for their comments. SGP acknowledges support from the Joint Committee ESO-Government of Chile. ULTRACAM, TRM, BTG, VSD and SPL are supported by the Science and Technology Facilities Council (STFC). The research leading to these results has received funding from the European Research Council under the European Union's Seventh Framework Programme (FP/2007-2013) / ERC Grant Agreement n. 267697 (WDTracer). MRS thanks for support from FONDECYT (1100782) and Millennium Science Initiative, Chilean ministry of Economy: Nucleus P10-022-F. The results presented in this paper are based on observations collected at the European Southern Observatory under programme IDs 087.D-0046 and 090.D-0277.

\bibliographystyle{mn_new}
\bibliography{eclipsers}

\label{lastpage}

\end{document}